\documentclass[11pt]{article}
\usepackage{epsfig} 
\usepackage{pifont}
\newcommand{\gtsim}{\raisebox{-0.5ex}{$\,\stackrel{>}{\sim}\,$}}
\newcommand{\ltsim}{\raisebox{-0.5ex}{$\,\stackrel{<}{\sim}\,$}}
\usepackage{cite}

\begin{document}

\renewcommand{\thefootnote}{\fnsymbol{footnote}}

\begin{center}
{\Large\bf Compton Scattering by the Proton}
\footnote[1]{Supported by the Italian Istituto
Nazionale di Fisica Nucleare (INFN) and by Deutsche
Forschungsgemeinschaft (SFB 201)
and by DFG-contracts Schu222 and 436RUS113/510}
\end{center}

\renewcommand{\thefootnote}{\arabic{footnote}}

\noindent
G. Galler$^a$, V. Lisin$^b$, R. Kondratiev$^b$, A.M. Massone$^c$,
S. Wolf$^a$, J. Ahrens$^d$, H.-J. Arends$^d$, R. Beck$^d$,
M. Camen$^a$,  G.P. Capitani$^e$, P. Grabmayr$^f$,
S.J. Hall$^g$,  F. H\"arter$^d$, T. Hehl$^f$, P. Jennewein$^d$,
K. Kossert$^a$, A.I. L'vov$^h$,  C. Molinari$^c$, P. Ottonello$^c$,
J. Peise$^d$, I. Preobrajenski$^d$, S. Proff$^a$, A. Robbiano$^c$,
M. Sanzone$^c$,
M. Schumacher$^{a,}$\footnote{corresponding author:
schumacher@physik2.uni-goettingen.de},
M. Schmitz$^d$,
F. Wissmann$^{c,}$\footnote{present address: Zweites Physikalisches Institut,
Universit\"at G\"ottingen}

\noindent
{\it $^a$ Zweites Physikalisches Institut, Universit\"at G\"ottingen,
D-37073 G\"ottingen, Germany}\\
{\it $^b$ Institute for Nuclear Research, Moscow 117312, Russia}\\
{\it $^c$ Dipartimento di Fisica dell'Universita di Genova and
INFN - Sezione di Genova, I-16146 Genova, Italy}\\
{\it $^d$ Institut f\"ur Kernphysik, Universit\"at Mainz, D-55099 Mainz,
Germany}\\
{\it $^e$ INFN - Laboratori Nazionali di Frascati, I-00044 Frascati, Italy}\\
{\it $^f$ Physikalisches Institut, Universit\"at T\"ubingen, D-72076
T\"ubingen, Germany}\\
{\it $^g$ Kelvin Laboratory, Glasgow University, Glasgow, Scotland,
G12 8QQ, UK}\\
{\it $^h$ P. N. Lebedev Physical Institute, Moscow 117924, Russia}\\

\begin{abstract}
Compton scattering by the proton has been measured over a wide range
covering photon energies
$250~{\rm MeV} \ltsim E_\gamma \ltsim 800~\rm MeV$
and photon scattering angles
$30^\circ \ltsim \theta_\gamma^{\rm lab} \ltsim 150^\circ$,
using the tagged-photon facility at MAMI (Mainz) and the
large-acceptance arrangement LARA. The data are in good agreement with
the dispersion theory based on the SAID-SM99K parameterization of
photo-meson amplitudes.
From the subset of data between 280 and 360 MeV the resonance
pion-photoproduction amplitudes were evaluated leading to
the multipole E2/M1 ratio
$\rm EMR(340~MeV)=(-1.6 \pm 0.4_{stat+syst} \pm 0.2_{model})\%$.
From all data below 455 MeV the proton's backward spin polarizability
was determined to be
$\rm \gamma_\pi =(-37.9 \pm 0.6_{stat+syst} \pm 3.5_{model})
 \times 10^{-4}~ fm^4$.

\end{abstract}


\noindent
$Keywords:$ Compton scattering, proton,
    scattering amplitudes, spin polarizabilities \\

\section{Introduction}

Compton scattering by the proton in the $\Delta$ energy-region has
played an important role in recent nucleon-structure investigations
carried out at the tagged-photon facilities at
Saskatoon (SAL) \cite{hallin93},
Brookhaven (LEGS) \cite{blanpied96,blanpied97,tonnison98}
and Mainz (MAMI) \cite{molinari96,peise96,huenger97,wissmann99}.
In addition, data are available for the second resonance region
measured in Tokyo \cite{ishii80,wada84} and Bonn \cite{jung81}
in the early 1980's.
Though important results have been obtained in these former experiments,
the experimental techniques had the disadvantage that only one angle
$\theta$ and a relatively small interval of photon energies $E_\gamma$
were available with a given configuration of the experimental set-up.
The present work reports about the first experiment where  large
ranges of $\theta$ and $E_\gamma$ were covered simultaneously through
the use of tagged photons and large acceptance arrangements for the
scattering angle and the photon energy.

The interpretation of the data obtained in the present experiment 
is facilitated by a recent  progress
\cite{lvov97,drechsel99} in the dispersion theory of proton Compton
scattering which is used as a precise tool for studying electromagnetic
properties of the nucleon. These properties include the electric and
magnetic polarizabilities $\alpha$ and $\beta$, the four spin
polarizabilities $\gamma_i$ with  the backward spin polarizability
$\gamma_\pi$ being a particular linear combination of them, and the
strength $M_{1+}$ and the multipole ratio E2/M1 of the $N\to\Delta$
transition. These quantities enter into the theoretical Compton
differential cross section as (not fully independent) parameters and
they are predominantly important in the $\Delta$ energy range.

Earlier versions of the dispersion theory of Compton scattering were
restricted to the first resonance region and failed at higher energies.
An extension of the dispersion theory to the second resonance region
was obtained only recently in \cite{lvov97}, using more accurate
photo-meson amplitudes and applying special measures to
suppress divergences of partial-wave series in dispersion integrals at
large $t$ (see \cite{lvov97} for details).
The  success of this  theory in wide energy and angular ranges is an
important prerequisite for its
reliable application in the $\Delta$-isobar region.

A more recent  version of the dispersion theory \cite{drechsel99}
improves on the previous work \cite{lvov97} by
considering  more quantitatively the  dynamics of the $t$-channel
two-pion exchange. In \cite{lvov97} this process was treated highly
phenomenologically by  approximating  asymptotic
contributions by effective-meson exchanges,  where the $\sigma$-exchange
in case of the invariant amplitude $A_1$ is of prominent importance.  
However, in its current version  the newer theory \cite{drechsel99}
unfortunately is less sophisticated in its treatment of 
pion-photoproduction contributions.
This makes the predictions of \cite{drechsel99} unstable in the region
$E_\gamma \gtsim 350$~MeV which contains 3/4 of the experimental data
obtained in the present experiment. For that reason we rely in our
analysis entirely on the theoretical framework of  \cite{lvov97}
and include uncertainties in the parameters of the asymptotic contributions
into model errors of the extracted physical quantities.

In comparison with the original version described in \cite{lvov97}, we
use a slightly updated code taking into account preliminary results
of the Mainz GDH experiment \cite{arends00} for  a more
accurate calculation of double-pion photoproduction contributions to the
dispersion integrals,  and taking into account the $\eta$, $\eta'$ 
$t-$channel exchanges with couplings borrowed from \cite{lvov99}.

\section{Experiment}

The present paper contains the results of an experiment carried out
using the {\bf LAR}ge Acceptance {\bf A}rrangement (LARA) shown in
Fig.~1.  This arrangement had been designed to cover the whole angular
range of photon scattering-angles from $\theta_\gamma^{\rm lab} = 30^\circ$ to
$150^\circ$ in the laboratory and the interval of photon energies from
$250 \; \mathrm{MeV}$ to $800 \; \mathrm{MeV}$. The principle of the
present method is to make use of the energy of the incident photon, of
the direction of the scattered photon and of the direction and energy
of the recoil proton to separate Compton scattering events from
background being mainly due to the ($\gamma,\pi^0$) reaction.

The experiment makes use of the Glasgow tagged photon facility
\cite{anthony91} installed at the $855 \; \mathrm{MeV}$ three-stage
microtron MAMI in Mainz \cite{herminghaus76}. The energy resolution
achieved by  the tagger was $\Delta E_\gamma = 2 \; \mathrm{MeV}$ on
the average.  The scattering target consists of lq.~$\mathrm{H}_2$
contained in a Kapton cylinder of $200 \; \mathrm{mm}$ length and
$30 \; \mathrm{mm}$ diameter.

On the photon arm 150 lead glass photon detectors (LG) were used having
dimensions of $15 \; \mathrm{cm} \times 15 \; \mathrm{cm} \times 30 \;
\mathrm{cm}$ positioned cylindrically around the scattering target with
the front faces having distances of $200 \; \mathrm{cm}$ from the
target center.  This leads to an angular resolution on the photon arm
of $\pm 2.2^\circ$ both in the horizontal and the vertical direction.
Each block containing 3 (horizontal) $\times$ 5 (vertical) detectors is
equipped with a plastic scintillator (VD) of $1 \; \mathrm{cm}$
thickness to identify charged background.

On the proton arm the proton angle $\theta_\mathrm{p}^{\rm lab}$ 
with respect to
the incident photon beam is  determined  by two wire chambers (WC) at
distances of $25 \; \mathrm{cm}$ and $50 \; \mathrm{cm}$ from the
target center. Each of these wire chambers consists of two layers of
wires tilted against the vertical direction by $+30^\circ$ and
$-30^\circ$, respectively.  The distance between wires in the layers is
2.5 mm.  The resolution achieved for the proton angle is better than
$1^\circ$ in the horizontal (geometrical $0.13^\circ$) and vertical
(geometrical $0.47^\circ$) directions. The proton energy is determined
via time-of-flight, measured through coincidences between signals from
the tagger and signals from 43 bars of $20 \; \mathrm{cm} \times 300 \;
\mathrm{cm} \times 5 \; \mathrm{cm}$ plastic scintillators (TOF)
\cite{grabmayr98}. The latter are arranged in 4 planes positioned at
distances of 2.6, 5.7, 9.4 and $12.0 \; \mathrm{m}$ from the target
center. The experiment trigger was defined through a coincidence
between a signal from a lead glass block and a signal from one out of 8
trigger detectors (TD) positioned behind the wire chambers, with the
geometry fulfilling the angular constraints of a Compton event.

\section{Data analysis}

Protons were identified through their comparatively large
energy-deposit in a  TD detector and through their time-of-flight.  For
each proton event detected by a TOF detector a trajectory was
constructed using the intersection points in the two wire chambers. The
event was accepted as a good one if the trajectory intersected the
scattering target, hit the appropriate TD detector and intersected the
TOF detector at the experimental impact point within its spacial
resolution.  Then, for a given proton trajectory and a given incident
photon energy $E_{\gamma}$ the direction of the scattered  photon
$\theta^{\rm Comp}_{\gamma}$ as well as the energy $E^{\rm Comp}_\mathrm{p}$
of the recoil proton were calculated assuming Compton kinematics.
Only those events were accepted where the experimental direction of the
secondary photon  was in agreement with the direction calculated  for a
Compton photon within the spacial resolution of the apparatus. This
procedure led to a drastic reduction of the number of background events
from $\pi^0$ photoproduction.

In addition to the separation procedures discussed above a further very
effective separation of events from Compton scattering and $\pi^0$
photoproduction was achieved by time-of-flight analysis.  The
experimental time-of-flight was compared with the one calculated from
the energy $E^{\rm Comp}_\mathrm{p}$ expected for a recoil proton of a
Compton event. Mean energy losses of the proton were used in this
calculation. The difference between the experimental and the calculated
time-of-flight was  named the missing time $\Delta t_{\rm p}$.

The analysis of the experimental data was accompanied by a Monte Carlo
simulation to determine the detector efficiencies. All calibrations
needed as inputs for a precise simulation, including the efficiencies
of the wire chambers,  were found in a calibration procedure making use
of the large amount of ($\gamma$,$\pi^0$) events provided by the
experiment.

Fig.~2 shows a  typical missing time  spectrum for primary photon
energies of $E_\gamma$ = 345 MeV. At this  low energy
there is a complete separation of the two types of events  whereas  at
higher energies in the first resonance region  there is some overlap
which can be removed by subtracting the tail of the ($\gamma$,$\pi^0$)
events underneath the Compton peak.

In the second resonance region it was necessary to make use of
$\pi^0$-background subtraction. 
The procedure is described in Fig. 3 for a photon energy 
of E$_\gamma$ = 780 MeV. 
Subfigure A shows Compton plus  background events, measured under conditions
where the trajectories of the incident photon, the recoil 
proton and the scattered photon 
were  located in one  plane within the spacial resolutions of the detectors. 
The variables $\cos_{\mbox{miss}}$ and E$_{\mbox{miss}}$ are the cosine of the
measured proton angle and the measured proton energy, respectively,
on a scale where these quantities are equal to zero for a Compton event.
Subfigure B shows background events to be subtracted from the data of
subfigure A. They are measured out-of-plane but otherwise under comparable
conditions. Any kinematical differences between the background events in A and
B are eliminated, by adjusting the data of B to the kinematical conditions of
A along the predictions of a  Monte Carlo simulation. It the
apparent that the vertical projection {\bf C} as well as the horizontal
projection {\bf E} of the data contained  in the rectangular boxes of
subfigures {\bf A} and {\bf B} lead to net Compton events of good
precision, as shown in the subfigures {\bf D} and {\bf F},
respectively.

\section{Results and Discussion}

Fig.~4 shows differential cross sections of the present experiment
for three scattering angles and energies in the first and second
resonance regions in comparison with data from earlier experiments.
Three theoretical curves were calculated using the SAID-SM99K
parameterization of the photo-meson amplitudes \cite{arndt96}
and three  different mass parameters $m_\sigma$ of the effective-$\sigma$
exchange. They demonstrate the sensitivity to $m_\sigma$ and show that
the choice made in \cite{lvov97}, i.e.  $m_\sigma \approx 600$ MeV,
works reasonably well.

Fig.~5 and 6  show further examples of differential cross sections
obtained for the first resonance region.  The errors shown contain all
contributions which are individual (random) for each data point and
have been carefully determined during the evaluation procedure. These
are the errors due to the counting statistics and the systematic errors
due to the detection efficiency, the geometrical uncertainty of the
apparatus and the background subtraction procedure. There  are
additional common (scale) systematic errors due to the tagging
efficiency ($\pm 2\,\%$) and target density and thickness ($\pm2\,\%$).
The two photon energies selected in Fig.~6 are chosen such that the
sensitivity of the quantities  $\gamma_\pi$ and E2/M1 to  the
differential cross sections is clearly demonstrated,  whereas for the
analysis described below a wider sample of our data has been taken into
account.  Keeping this in mind, the fluctuations of some of the data
shown in Fig.~6 are not indicative of a lack of  precision of the results.
Fig.~5 clearly proves that rather precise information on the properties
of $p \to \Delta$ transition may be obtained from the total amount of
data obtained in the first resonance region.

In order to determine the multipoles  characterizing  
the $\Delta$-resonance and to
extract $\gamma_\pi$ we use the following procedure.  We start with the
fixed  mass parameter $m_\sigma=600$ MeV and the difference of the
electric and magnetic polarizabilities of the proton,
$\alpha-\beta=10.0\times 10^{-4}~\rm fm^3$, as determined from
low-energy Compton scattering experiments \cite{macgibbon95}.  Taking a
subset of 167 data points close to the $\Delta$-resonance peak, namely those
between the limits $E_\gamma = 280$ and 360 MeV where the
$\Delta$-resonance contribution strongly dominates, we slightly rescale the
$\Delta$-resonance parts of the photo-pion amplitudes $M_{1+}$ and
$E_{1+}$, as described in \cite{huenger97}, in order to achieve the
best agreement between the present experimental data and dispersion-theory
predictions. The above
choice of the energy limits is made in order to reduce otherwise bigger
model errors in the determination of the resonance parameters.
With these corrected amplitudes, setting an overall scale for the
theoretical differential cross sections of Compton scattering close to
the resonance, we tune $\gamma_\pi$ through the asymptotic contribution
to the invariant amplitude $A_2$ (cf.\ \cite{tonnison98}) in order to
arrive at the best $\chi^2$ in the whole energy region covering the
$\Delta$-resonance, which here is the region $E_\gamma \le 455$ MeV
containing 467 data points.
With this $\gamma_\pi$ we repeat the determination of the amplitudes
$M_{1+}$ and $E_{1+}$
and then arrive  again at $\gamma_\pi$, etc. These iterations quickly converge 
and eventually give the final values  for $M_{1+}$, $E_{1+}$ and
$\gamma_\pi$.

In order to determine the  model uncertainties of  the extracted quantities
we used different values for $\alpha-\beta$ within the experimental 
uncertainty of this quantity (i.e.\
between 8.5 and $11.5\times 10^{-4}~\rm fm^3$
\cite{macgibbon95,baranov00}).
Also different values for $m_\sigma$ were used between 500 to 700 MeV. 
This range of $m_\sigma$ is supported by a comparison of different theoretical
calculations of the amplitude $A_1$ 
\cite{lvov97,drechsel99,holstein94,lvov99a}.
Moreover, we varied the $\pi^0\gamma\gamma$ coupling by $\pm$4\% 
and the $\eta NN$ and $\eta'NN$ couplings by $\pm$50\%.
The form factors accompanying the $\pi^0$, $\eta$, $\eta'$ t-channel
contributions were varied and also the parameters
which determine the multipole structure
of double-pion photoproduction below 800 MeV where the latter variation
was based on experience of a recent GDH experiment \cite{arends00}.

We present our findings in terms of the absolute value of the
$M_{1+}^{(3/2)}$ amplitude at the energy 320.0 MeV corresponding to the
maximum of  the differential cross section for  Compton scattering.
The E2/M1 ratio (EMR) of the imaginary parts of the amplitudes
$E_{1+}^{(3/2)}$ and $M_{1+}^{(3/2)}$  is determined
for  340.0 MeV where the real parts of these amplitudes are about
zero, in complete agreement with the previous procedure  
\cite{blanpied97,beck97,beck99} where 
 the ratio of the imaginary parts was determined from pion
photoproduction experiments. It is important to exactly 
use the same energy $E_\gamma$ when comparing
the amplitudes $E_{1+}^{(3/2)}$ and $M_{1+}^{(3/2)}$ obtained 
from different experiments because they rapidly vary with  $E_\gamma$.
Our results are
\begin{eqnarray}
    |M_{1+}^{(3/2)}(320~{\rm MeV})|
      &=& (39.7 \pm 0.3_{\rm stat+syst} \pm 0.03_{\rm model})
        \times 10^{-3} / m_{\pi^+} ,
\nonumber \\
     {\rm EMR}(340~{\rm MeV})
      &=& (-1.6 \pm 0.4_{\rm stat+syst} \pm 0.2_{\rm model})\;\% ,
\nonumber \\
     \gamma_\pi
      &=& (-37.9 \pm 0.6_{\rm stat+syst} \pm 3.5_{\rm model})
        \times 10^{-4} ~\rm fm^4.
\end{eqnarray}
The systematic errors given here include changes imposed by 
a simultaneous shift
of all data points within the scale uncertainty of $\pm$3\%. This
uncertainty fully dominates the resulting uncertainty of  the
$M_{1+}^{(3/2)}$ amplitude.
Note that the required modifications of the  amplitudes $M_{1+}^{(3/2)}$
and $E_{1+}^{(3/2)}$ are 
compatible with zero. Without the modification,
the SAID-SM99K parameterization gives
$|M_{1+}^{(3/2)}(320~{\rm MeV})| = 39.74$ (in the same units) and
EMR$(340~{\rm MeV}) = -1.68\%$.
The present value for  $M_{1+}^{(3/2)}$ perfectly agrees with the one
previously determined by H\"unger et al.
\cite{huenger97}:  $|M_{1+}^{(3/2)}(320~{\rm MeV})| = 39.6 \pm 0.4$.
The value of EMR determined from the present Compton scattering data
is  smaller than the one  obtained in a dedicated Mainz  photo-pion
experiment, i.e. $(-2.5\pm 0.1_{\rm stat} \pm 0.2_{\rm syst})\%$ 
\cite{beck97,beck99},
and also smaller than the result published by the LEGS group
\cite{blanpied97}, i.e.
$(-3.0\pm 0.3_{\rm stat+syst}\pm 0.2_{\rm model})\%$.

The uncertainties of the spin polarizability $\gamma_\pi$
are dominated by the model errors, especially by the variations of
$m_\sigma$ and $\alpha-\beta$. Our result for $\gamma_\pi$ is in
disagreement with the one determined in 1997
by the LEGS group \cite{tonnison98} which gave the  smaller value
$\gamma_\pi^{\rm LEGS} = -27.1 \pm 2.2_{\rm stat+syst}{
^{+2.8}_{-2.4}}{}_{\rm model}$ (in the same
units of $10^{-4}~\rm fm^4$). This difference can be traced back to a
difference in the measured differential cross sections, as can be seen 
in Fig.~6.

The present value of $\gamma_\pi \approx -37.9$ agrees well with
predictions of the unsubtracted dispersion relation for the invariant
amplitude $A_2$ adopted in \cite{lvov97}. The latter gives $-38.24$
with the same photo-meson input and with the same
energy cut in the dispersion integrals of  $E_{\rm max}=1.5$ GeV,
thus assuming no essential asymptotic
contributions beyond pseudoscalar-meson exchanges ($\pi^0,\eta,\eta'$).
The present value for  $\gamma_\pi$ satisfactorily agrees with predictions
of the ``small scale expansion" scheme, which effectively is
chiral perturbation theory including the $\Delta$-resonance,
$\gamma^{\rm SSE}_\pi = -37$ \cite{hemmert98}. It also agrees with standard
chiral perturbation theory to order $O(p^4)$, which does not include the
$\Delta$-resonance, $\gamma^{\rm ChPT}_\pi = -39$ \cite{kumar00},
provided $-45$ is used for the anomaly contribution to $\gamma_\pi$
from $\pi^0$ exchange\footnote
{We do not use another ChPT prediction,
$\gamma^{\rm ChPT}_\pi = -42$ \cite{gellas00} for reasons
explained in \cite{birse00}.}. Furthermore, it agrees
with backward-angle dispersion relations, which include the
$\Delta$ and the $\eta$-$\eta'$ exchanges,
$\gamma^{\rm DR}_\pi = -39.5 \pm 2.4$ \cite{lvov99}.
Thus, there is good overall consistency between the present
Compton scattering data, the dispersion theory,
and the SAID-SM99K photo-meson amplitudes.

Such a consistency is deteriorated when the latest SAID-SM00K
photo-pion amplitudes are used. This is because in that latest
parameterization the $M1$-strength of the $\Delta$-resonance
is decreased to $|M_{1+}^{(3/2)}(320~{\rm MeV})| = 39.16$.
Therefore, we have to increase the SM00K $M_{1+}{(3/2)}$-amplitude 
by $+1.2\%$
in order to achieve  a satisfactory description of Compton scattering.
When such a rearrangement is made, the value extracted for $\gamma_\pi$ 
is $\gamma_\pi = - 37.7$, i.e. it turns out to be only slightly
smaller than the one of  Eq.(1) with similar errors.

When using the  MAID2K \cite{maid} parameterization of
photo-pion amplitudes
the same procedure gives the results
\begin{eqnarray}
    |M_{1+}^{(3/2)}(320~{\rm MeV})|
      &=& (39.8 \pm 0.3_{\rm stat+syst} \pm 0.03_{\rm model})
        \times 10^{-3} / m_{\pi^+} ,
\nonumber \\
     {\rm EMR}(340~{\rm MeV})
      &=& (-1.9 \pm 0.4_{\rm stat+syst} \pm 0.2_{\rm model})\;\% ,
\nonumber \\
     \gamma_\pi
      &=& (-41.4 \pm 0.4_{\rm stat+syst} \pm 2.5_{\rm model})
        \times 10^{-4} ~\rm fm^4
\end{eqnarray}
which are more at variance with Eq.~(1) than the alternatives discussed above.
In this case a slightly bigger rearrangement of the resonance
amplitudes is required in comparison with their original values
which, for MAID2K, are $|M_{1+}^{(3/2)}(320~{\rm MeV})| = 39.92$
and EMR$(340~{\rm MeV}) = -2.19\%$.
The biggest change is, however, in the spin polarizability
$\gamma_\pi$ which can be traced back to rather different
nonresonant amplitudes $E_{0+}$ and $E_{2-}$ in the SAID and MAID
representations in the $\Delta$-resonance range.
The overall quality of the description of the present Compton scattering data
at energies below 455 MeV,
containing 467 data points in total,
is approximately the same for the SAID and MAID photo-meson input.
The fitting procedure based on the two sets of parameterizations leads to
$\chi^2=564$ and 565, respectively, and the differences in the
predictions are small  as can be seen in  Fig.~5. 

However, the properties of the SAID and MAID parameterizations are quite
different in the second resonance region.  
For instance, $\chi^2=243$ is obtained for 190 data points 
between 450 and 600 MeV for SAID-based theoretical predictions with SAID-based
parameters (1), whereas $\chi^2=513$ is obtained for the same 190 data points
with MAID-based theoretical predictions and MAID-based parameters (2).
This means that the MAID-based parameterization does not lead to a reasonable
fit to the data when the same parameter $m_\sigma=600$ MeV is used.
The biggest difference between these two versions is seen at backward
angles in the dip region between the first and second nucleon resonance,
as illustrated by the dashed-dotted curve in Fig. 4.
The use of a smaller $m_\sigma$ with the {\em same} $\gamma_\pi$ 
reduces  the  discrepancy in the dip region, however without leading to an
overall agreement. It is observed that
the fit to the data  below 455 MeV carried out with that smaller
$m_\sigma$ requires an even bigger $-\gamma_\pi$ compared to the one given in
(2), and with this bigger
$-\gamma_\pi$ again no agreement is achieved between the theory and the data
in the dip region.

\section{Conclusions}
The results of the present experiment may be summarized as follows.
For the first time Compton scattering by the proton has been measured
with a large acceptance set-up for the scattering angle and the photon energy.
The data confirm the magnitude of the $M1$-strength adopted
in the SAID-SM99K and MAID2K parameterizations (not in SAID-SM00K),
and are in agreement with the E2/M1 ratio given by these
parameterizations.
The backward spin polarizability $\gamma_\pi$ is found
to be in agreement with latest theoretical calculations,
although model errors should yet be better understood.

\section*{Acknowledgment}
The authors are greatly indebted to Professor Turleiv Buran,
Department of Physics,
University of Oslo and to the Norwegian Research Council for Science and
the Humanities for having given them the opportunity to use their
120 lead glass detectors for this experiment,  containing
150 lead glass detectors in total.
They wish to acknowledge the excellent support of the accelerator group
of MAMI, as well as many other scientists and technicians of the Institut f\"ur
Kernphysik at the University of Mainz.

\newpage
\begin{figure}
\centering\epsfig{figure=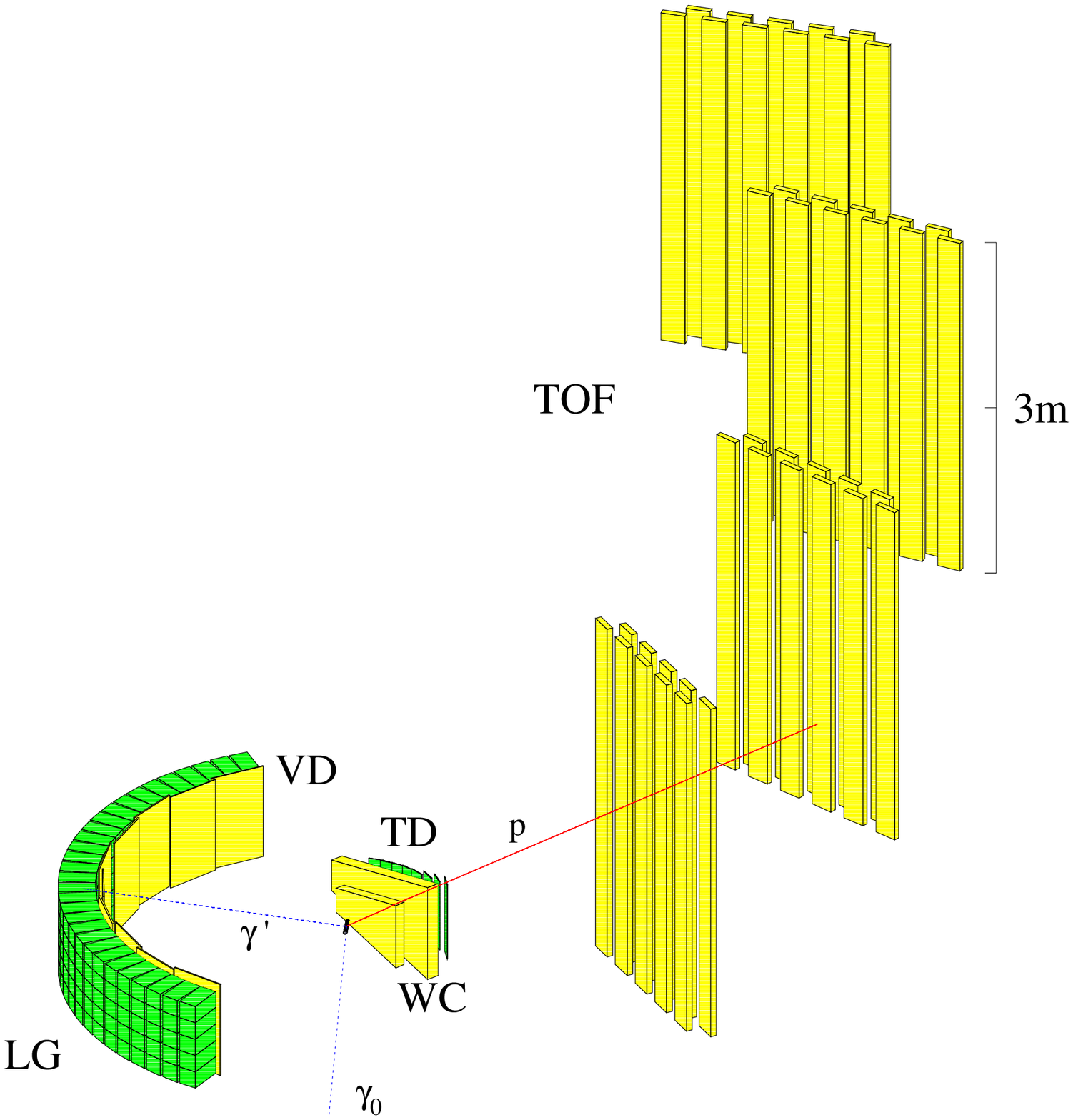,width=0.7\linewidth}
\caption{Perspective view of the LARA arrangement. Left: Photon arm
consisting of 10 blocks \`a 3 $\times$ 5 lead glass detectors (LG) each
equipped  with a 1~cm  plastic scintillator (VD). Right: Proton arm
consisting of two wire chambers (WC) at distances of 25 and 50 cm from the
target center, 8 plastic scintillators serving as trigger detectors (TD) and
43 bars of 20 cm $\times$ 300 cm $\times$ 5 cm plastic scintillators
serving as time-of-flight (TOF) stop detectors. The scattering target
consisted of lq. $\mathrm{H}_2$ contained in a 3 cm $\O$ $\times$ 20 cm Kapton
cylinder.}
\label{fig1}
\end{figure}

\newpage
\begin{figure}
\centering\epsfig{figure=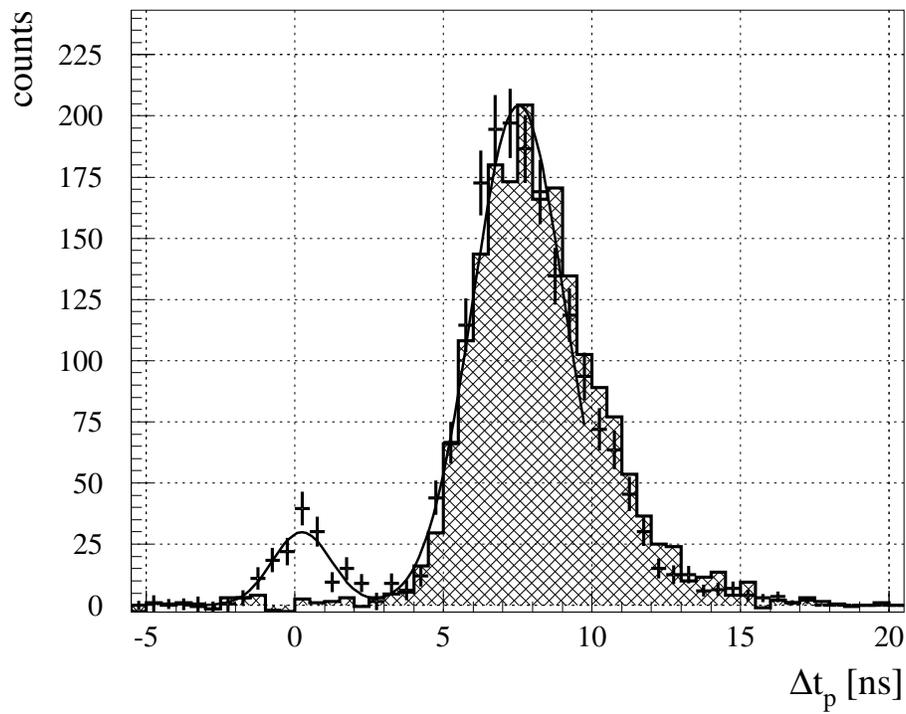,width=0.7\linewidth}
\caption{Missing time spectrum for  protons at  an incident  photon
energy of $E_\gamma = 345$ MeV, measured at a photon angle of
$\theta^{\rm lab}_{\gamma}$ = 70$^\circ$.
Left distribution:
($\gamma,\gamma$) events. Right (cross-hatched) distribution:
($\gamma,\pi^0$) events.}
\label{fig2}
\end{figure}

\newpage %%%%%%%%%%%%%new figure 3
\begin{figure}
\centering\epsfig{figure=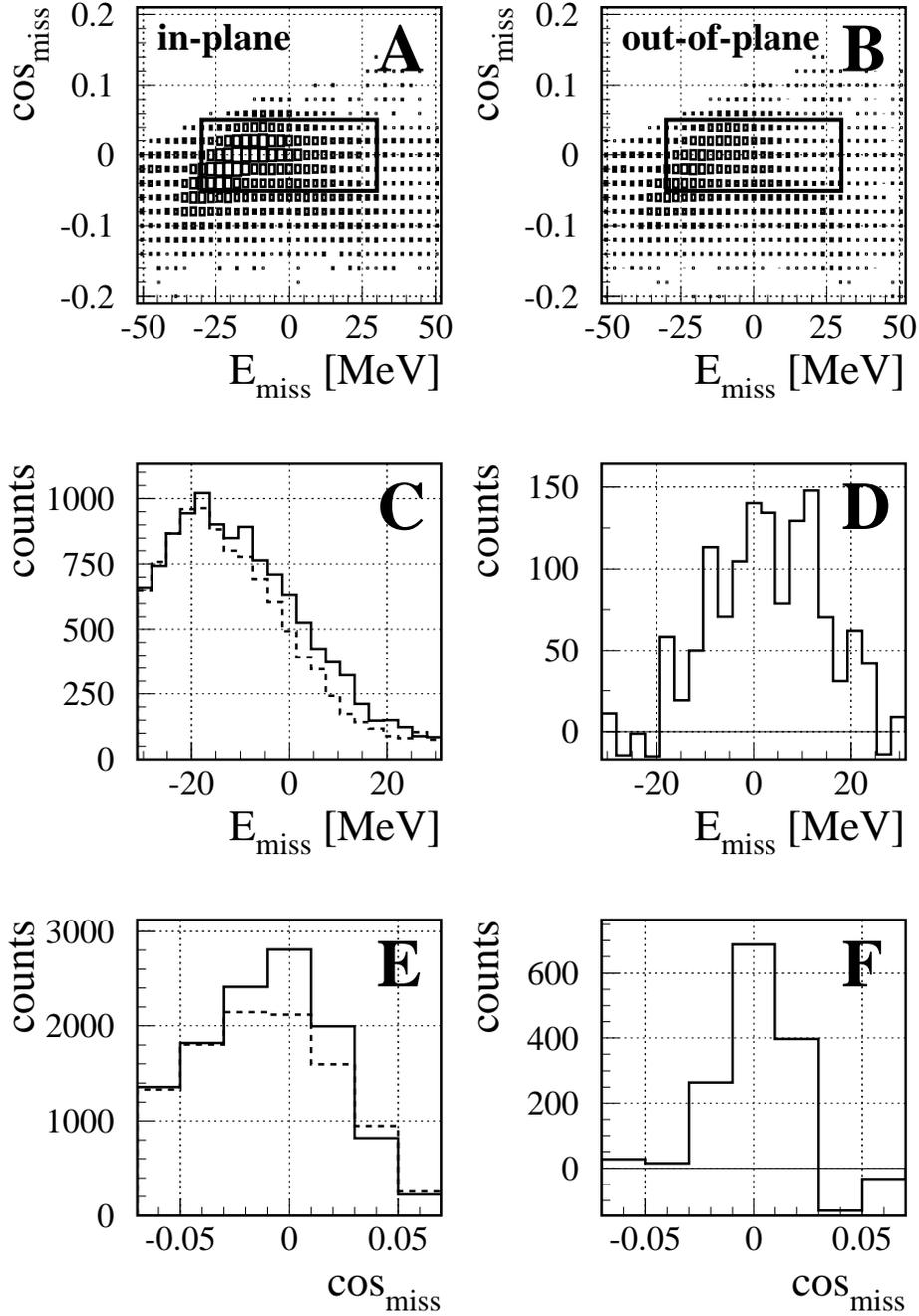,width=0.7\linewidth}
\caption{
Experimental data obtained for a photon energy of E$_\gamma$ = 780 MeV
at a scattering angle of $\theta^{\rm lab}_{\gamma}$ = 37$^\circ$.
{\bf A,B}: Scatter  plots of events measured inside and outside
the Compton scattering plane, with $\cos_{\rm miss}$ and E$_{\rm miss}$ being
the missing cosine and missing energy of the proton, respectively,
with respect to a
Compton event. {\bf C,E}: Vertical and horizontal projections, respectively,
of data inside the rectangular boxes of {\bf A} (solid) 
and {\bf B} (dashed). {\bf D,F}:
Net ($\gamma,\gamma$) events from
 {\bf C} and {\bf E}, respectively.
}
\label{fig3}
\end{figure}

\newpage
\begin{figure}
\centering\epsfig{figure=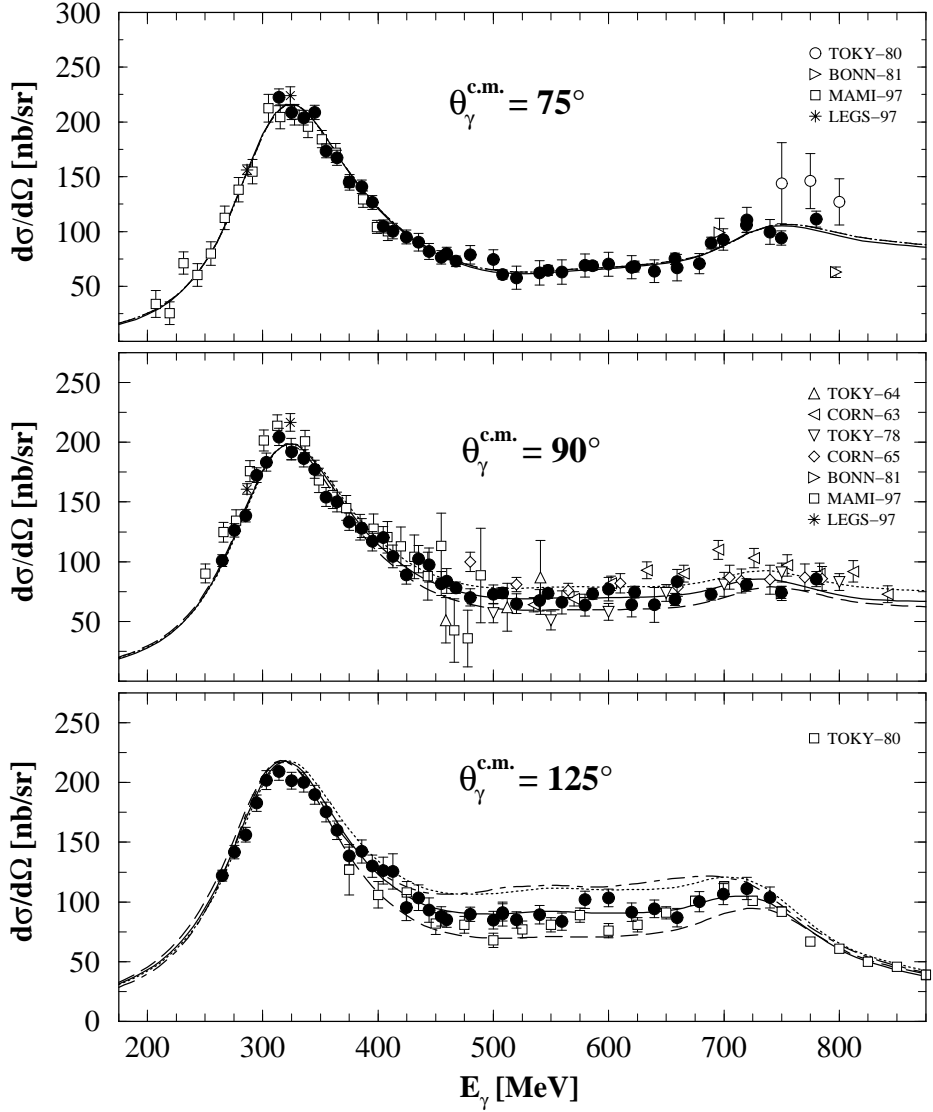,width=0.7\linewidth}
\caption{Differential cross sections for the first and second resonance region
in comparison with data from other experiments. The curves show calculations
based on the SAID-SM99K photo-meson amplitude
for $m_\sigma = 400 \; \mathrm{MeV}$ (dashed),
$m_\sigma = 600 \; \mathrm{MeV}$ (solid)
and $m_\sigma = 800 \; \mathrm{MeV}$ (dotted).
Other parameters are those in Eq.~(1).
The dashed-dotted curve given for the angle $125^\circ$
shows calculations based on the MAID2K photo-meson
amplitudes with $m_\sigma=600$ MeV and other parameters
specified in Eq.~(2).
The previous data are
compiled in \cite{ukai97} and are taken from:
\cite{ishii80} (TOKY-80);
\cite{jung81} (BONN-81);
\cite{huenger97} (MAMI-97);
\cite{tonnison98} (LEGS-97);
\cite{nagashima64} (TOKY-64);
\cite{stiening63} (CORN-63);
\cite{toshioka78} (TOKY-78);
\cite{rust65} (CORN-65).
The data of the present work (filled circles, representing  angular intervals
of $\Delta \theta^{\,\rm c.m.}_\gamma = 15^\circ$) are given
with error bars taking into account the counting statistics, 
and systematic errors due
to detection efficiency, geometrical uncertainties and background subtraction.}
\label{fig4}
\end{figure}

\newpage
\begin{figure}
\centering\epsfig{figure=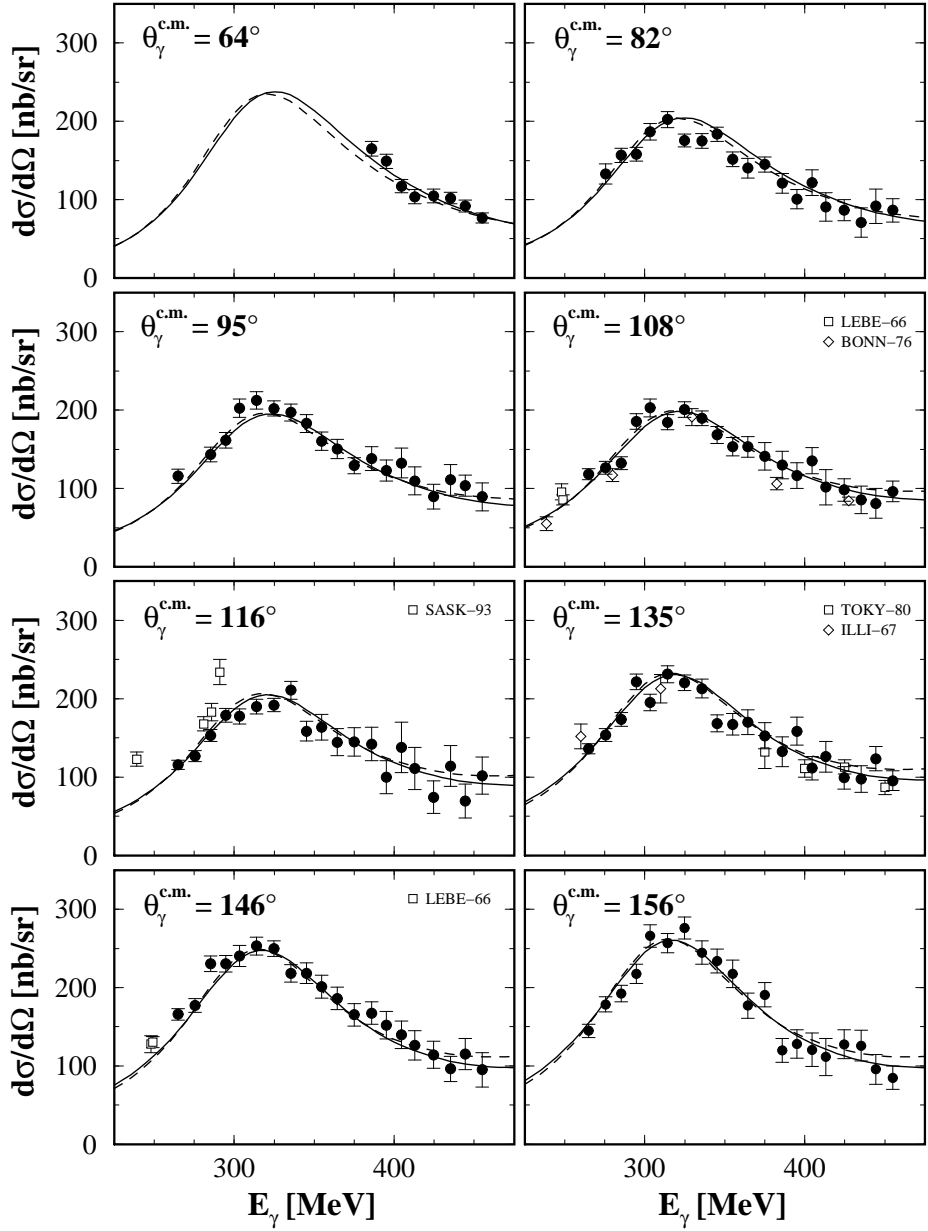,width=0.7\linewidth}
%\centering\epsfig{figure=LaraPics/energieverteilung.eps,width=0.7\linewidth}
\caption{8 out of 24 energy distributions from 59$^\circ$
to 156$^\circ$ (c.m.) obtained with the LARA arrangement in the first
resonance range compared with previous data and with predictions from
dispersion theory
(SAID-SM99K --- solid lines, MAID2K --- dashed lines).
The previous data are taken from:
%\cite{bernardini60} (ILLI-60);
%\cite{dewire61} (CORN-61);
%\cite{nagashima64} (TOKY-64);
\cite{baranov66a, baranov66b} (LEBE-66);
\cite{gray67} (ILLI-67);
\cite{genzel76} (BONN-76);
\cite{hallin93} (SASK-93);
%\cite{blanpied96} (LEGS-96);
%\cite{huenger97} (MAMI-97);
\cite{ishii80} (TOKY-80).
The data of the present work (filled circles, representing 
angular intervals of $\Delta\theta_\gamma^{\,\rm c.m.}=4^\circ$) 
are given with error bars as in Fig. 4.}
\label{fig5}
\end{figure}

\newpage
\begin{figure}
\centering\epsfig{figure=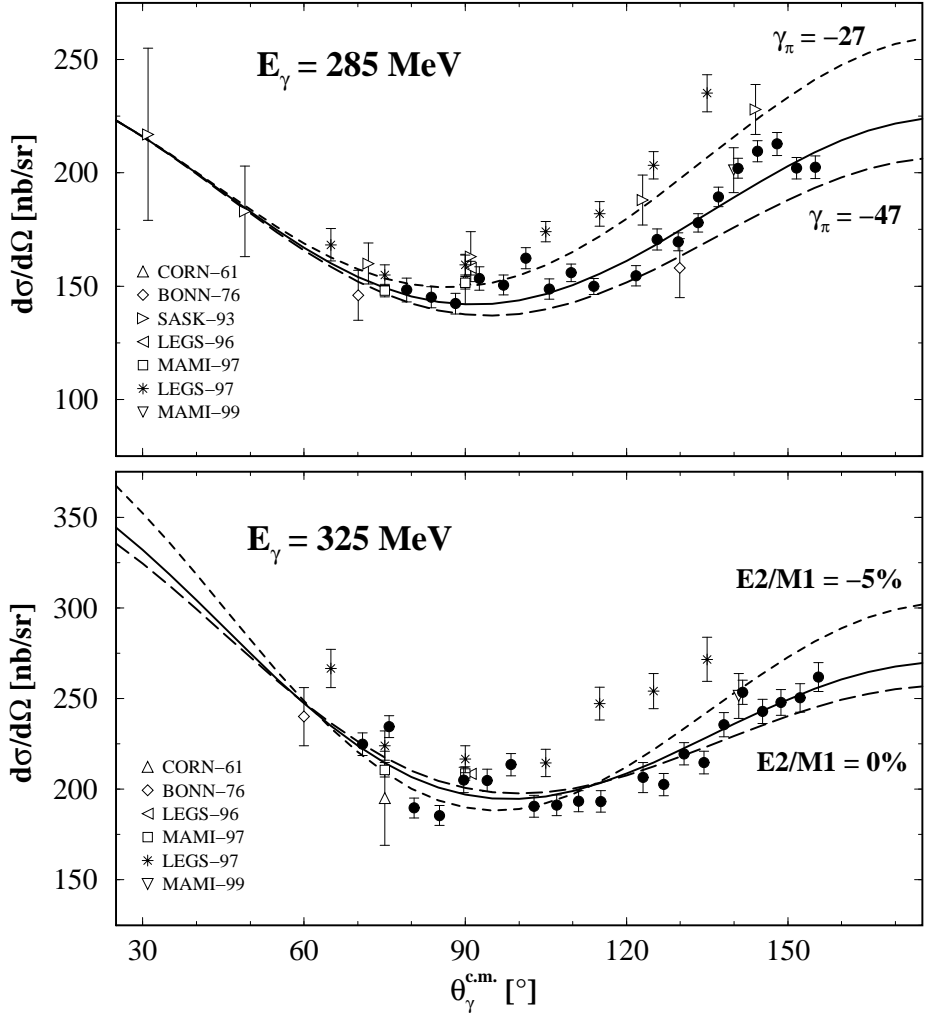,
width=0.7\linewidth}
\caption{Angular distributions of Compton differential cross sections
obtained with the LARA arrangement (filled  circles,
representing energy intervals of $\Delta E_\gamma=40$ MeV) 
compared with previous data
as compiled in \cite{ukai97} and with predictions of dispersion
theory with the SAID-SM99K photo-pion amplitudes.
The standard parameterization is given by the full line (SAID-SM99K,
$\gamma_\pi = -37.9 \times 10^{-4} \; \mathrm{fm}^4$).
The dashed lines show sensitivities to $\gamma_\pi$ at
$E_\gamma = 285 \; \mathrm{MeV}$ (upper part) and to the ratio E2/M1 at
$E_\gamma = 325 \; \mathrm{MeV}$ (lower part).
The previous data are from: \cite{dewire61} (CORN-61); \cite{hallin93}
(SASK-93); \cite{genzel76} (BONN-76); \cite{blanpied96} (LEGS-96);
\cite{huenger97} (MAMI-97); \cite{tonnison98} (LEGS-97); \cite{wissmann99}
(MAMI-99). The final value for the parameter $\gamma_\pi$ has
not been obtained
from these data points only but from the total amount of
data available below 455 MeV (see text).}
\label{fig6}
\end{figure}

\end{document}